\documentstyle[twocolumn,prb,aps,epsf]{revtex}

\begin{document}

\draft

\title{
Critical Temperature $T_{c}$ and Charging Energy $E_{c}$
between B-B layers of
Superconducting diboride materials MgB$_{2}$  in 3D JJA model
}

\author{
Chikao Kawabata,$^{1}$
Nobuhiko Hayashi,$^{2}$
and Fumihisa Ono$^{3}$
}

\address{
$^{1}$Division of Liberal Arts and Sciences,
$^{2}$Computer Center,
$^{3}$Department of Physics,\\
Okayama University, 2-1-1 Tsushima-Naka, Okayama 700-8530, Japan
}
\date{\today}
\maketitle

\begin{abstract}
   The diboride materials MB$_{2}$
(M$=$ Mg, Be, Pb, etc.)
are discussed
on the basis of the 3D Josephson junction array (JJA) model
due to Kawabata-Shenoy-Bishop,
in terms of
the B-B layers in the diborides analogous to the Cu-O
ones in the cuprates.
   We propose a possibility of
superconducting materials
with the MgB$_{2}$-type structure
which exhibit higher critical temperature $T_c$ over 39K of MgB$_{2}$.
   We point out a role of
interstitial ionic atoms (e.g., Mg in MgB$_{2}$)
as capacitors between the B-B layers,
which reduce the charging coupling energy
in JJA.
\end{abstract}

\pacs{KEYWORDS:\quad
MgB$_{2}$, Josephson Junction Array model,
Ionic electronic polarizability, Charging energy}

   Since Akimitsu and co-workers presented a remarkably high
temperature ($\simeq$ 39K)
superconducting magnesium diboride material MgB$_{2}$
in the beginning of 2001,~\cite{nagamatsu01}
much attention
has been focused on studying the diboride material
superconductivity.
\cite{bianconi01,kortus01,an01,satta01,mehl01,jin01,hirsch01,felner01}
   The material MgB$_{2}$
takes the layered structure of B-Mg-B stacking and has a coupling between
2-dimensional (2D)
B-B layers.~\cite{nagamatsu01,bianconi01}
   It is believed from several band calculations, that
in the MgB$_{2}$-type crystal structure
there exist a strong covalent B-B bonding
and an ionic B-Mg bonding.~\cite{kortus01,an01,satta01,mehl01}
   The strong covalent B-B bonding and the presence of the ionic Mg atoms
lead to hole-like
cylindrical Fermi surfaces in MgB$_{2}$,~\cite{kortus01,an01}
and the holes are conducted in each 2D B-B layer.
   It is observed that
the 2D B-B layers,
like the Cu-O layers in high-$T_c$ cuprates, play an important role
in the electrical transport properties.~\cite{jin01}
   MgB$_{2}$ can then be beautiful realization of the essential physics of
superconductivity in cuprates, without the complications of
Cu $d_{x^2-y^2}$ orbitals.~\cite{hirsch01}

   In this short note, in terms of a theoretical model
of the B-B layers analogous to the Cu-O
layers,
we propose a possibility of
superconducting materials
of the MgB$_{2}$-type structure
with higher critical temperature $T_c$ over 39K.
   We point out an important role of
interstitial ionic atoms (e.g., Mg in MgB$_{2}$)
as capacitors between the layers.

   We model the diboride MB$_{2}$ materials
(M$=$ Mg, Be, etc.)
as quantum-capacitive
Josephson junction arrays (JJA)
with weakly coupled superconducting grains
on a 3D lattice and angstrom-scale parameters,
following Kawabata-Shenoy-Bishop
(hereafter, we call KSB)
theory,~\cite{kawabata94a,kawabata95a,kawabata95b,kawabata94b,kawabata96}
originally proposed to the high-$T_c$ cuprates
with 2D Cu-O layers and a coupling between the Cu-O
layers.
   It means that the Josephson junction coupling energy $E_{j}$ via
the Cooper pair tunneling within and between the B-B layers enhances
the critical temperature $T_{c}$ for layered B-B superconductors,
while the charging coupling energy $E_{c}$
between the B-B layers
{\it depresses} $T_{c}$.
The origin of $E_{c}$ is that,
when the Cooper pairs tunnel
between the layers, local inter-layer charge unbalance is induced
between the superconducting grains in JJA.
   Here, the interstitial ionic atoms, M,
located between the B-B layers play a role of
capacitors,
reducing the charging coupling energy $E_{c}$
by the electric polarization of the ionic atom.

   Referring to Kittel's textbook and
Shockley's
tables~\cite{kittel,tessman53,pauling27,pearson}
for the ionic radii $d$ and electronic
polarizability $\varepsilon$ of the ionic atoms M,
we have obtained the charging coupling energy
$E_{c} \sim d / \varepsilon$
between the B-B layers in Table~\ref{table:Ec}, following
the KSB theory.
\cite{kawabata94a,kawabata95a,kawabata95b,kawabata94b,kawabata96}
   By hypothesizing same
Josephson junction coupling
energy $E_{j}$ within the B-B layers for
any diboride materials,
we estimate critical temperatures
$T_{c}$ ($=f(E_c)$)
considering the competition between
the Josephson junction coupling energy $E_{j}$ and
the charging coupling energy
$E_{c}$~\cite{kawabata94a,kawabata95a,kawabata95b,kawabata94b,kawabata96}
for the diboride compounds
MB$_2$.
   We here discuss, by way of illustration,
the ionic compounds M$^{2+}$(B$^{-}$)$_2$
expected to have nearly the same electronic structure~\cite{satta01}
as MgB$_{2}$ (Mg$^{2+}$(B$^{-}$)$_2$),
and then to have almost the same $E_{j}$ (i.e., a ``bare'' $T_c$
without any influence of
the M-atom-dependent charging coupling energy $E_c$).

\begin{figure}
\epsfxsize=80mm
\begin{center}
\epsfbox{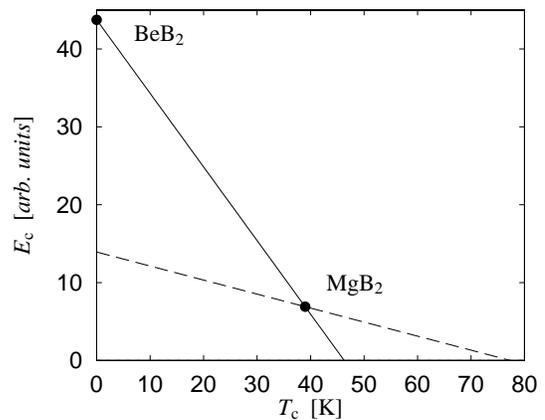}
\end{center}
\caption{
   Charging coupling energy $E_c$ versus critical temperature $T_c$.
   Points are for MgB$_2$ ($T_c = 39$K ~\cite{nagamatsu01})
and BeB$_2$ ($T_c \simeq 0$K ($<5$K) ~\cite{felner01}).
   The values of $E_c$ are of Table~\ref{table:Ec}
for Mg$^{2+}$ and Be$^{2+}$.
   Following these lines on the $E_c$ vs.\ $T_c$ plot (see text),
compounds containing ionic atoms M with smaller $E_c$
than Mg$^{2+}$ in Table~\ref{table:Ec},
are predicted to exhibit higher $T_c$
than 39K of MgB$_2$.
}
\label{fig:tc}
\end{figure}

   In Fig.~\ref{fig:tc} for the $E_{c}$ vs.\ $T_{c}$ plot,
we show two points,
as experimentally known data,
for BeB$_{2}$ ($T_{c} \simeq 0$K ($<5$K),~\cite{felner01} $E_{c}=43.75$)
and MgB$_{2}$ ($T_{c}=$39K,~\cite{nagamatsu01} $E_{c}=6.91$),
where $E_{c}$ ($\sim d/\varepsilon$) are of Table~\ref{table:Ec}
and in arbitrary units.
   These two are, at present, the only reported materials
with the MgB$_{2}$-type structure
on which $T_c$ are measured
experimentally.
   While only two points have been determined so,
we draw a straight line (solid line) as is shown
in Fig.~\ref{fig:tc}, assuming
simple linear relation, $T_{c} = f(E_c) \sim E_{j} - E_{c}$.
   We see that, when this solid line crosses the $T_{c}$-horizontal axis,
such a $T_{c}$-cross-point seems to be larger than 39K.
   For various dibrode compounds with different $E_c$
in Table~\ref{table:Ec},
data points align on the line in Fig.~\ref{fig:tc},
and then $T_c$ can be estimated for each compound.
   From this graph,
for compounds which have smaller $E_c$ than MgB$_2$,
$T_c$ higher than 39K of MgB$_2$ are obtained.
   That is, since $E_c$ of the ionic atoms in Table~\ref{table:Ec},
except for Be$^{2+}$,
are smaller than that of Mg$^{2+}$,
higher $T_c$ is estimated for various MB$_2$ compounds
owing to
the relation $E_c = f^{-1}(T_{c})$ as shown by the line in Fig.~\ref{fig:tc}.
   Especially for the compound containig the Pb atom,
high $T_c$ is expected because of
large poralizability $\varepsilon$ (thus, small $E_c$)
of the ionic Pb
in Table~\ref{table:Ec}.

   We represent, in what follows, two kinds of possibility that
$T_c$ may become higher than those estimated from the solid line
drawn in Fig.~\ref{fig:tc}.
   One possibility is that, if $E_{j} - E_{c} < 0$
for BeB$_{2}$,
which means it never exhibits superconductivity, i.e., $T_c =0$,
then we infer that the $E_{c}$ vs.\ $T_c$ plot becomes
a line with a smaller slope
as shown by the dashed line in Fig.~\ref{fig:tc}
because a (virtual) $T_c$ ($\sim E_{j} - E_{c}$)
for BeB$_{2}$
is negative
on the graph of Fig.~\ref{fig:tc} in this case.
   The other is that, while we have assumed the function $f(E_c)$
(or $f^{-1}(T_c)$) as
a linear line, it can be a curve with positive curvature
in general.~\cite{kawabata94a,kawabata95a,kawabata95b}
   A curve, $E_c = f^{-1}(T_c)$, with positive curvature
which goes through both the points
for MgB$_2$ and BeB$_2$ in Fig.~\ref{fig:tc},
could give rise to rather higher $T_c$ in the region of small $E_c$.
   We then expect that the critical temperatures $T_{c}$
could be possible to reach
up to the region over 77K, the evaporation temperature of liquid nitrogen.

   In conclusion, we indicate that, if MgB$_{2}$-type compounds are
synthesized by using ionic atoms with larger poralizability or
smaller diameter
instead of the Mg atoms, the critical temperatures $T_c$ would become higher.
   We hope that
the present proposal is an encouragement to the synthesis of compounds
and
new high-$T_c$ superconducting systems
over 39K
with the MgB$_{2}$-type structure
will be discovered in the near future.

   One of the authors (C.K.) would like to thank
Professor S.\ R.\ Shenoy,
Dr.\ A.\ R.\ Bishop
and
Dr.\ N.\ L.\ Saini
for discussions and encouragement
in the initial stage of this study.

\begin{table}
\caption{
   Ionic radii $d$ (from ref.\ 15),
ionic electronic polarizability $\varepsilon$
(from ref.\ 16)
and charging coupling energy $E_c \sim d/\varepsilon$.
   The data of $\varepsilon$ in brackets are
of ref.\ 17
originally
and the value of $d$ for Cu$^{2+}$ is of ref.\ 18.
}
\vspace{3mm}
\begin{tabular}{lllll}
Ionic &
Ionic &
Ionic &
Charging &
Critical \\
atom &  radii~\cite{kittel,pearson} &
 electronic &
 coupling &
temperature \\
 & &
poralizability~\cite{tessman53,pauling27} &
energy &
 \\
M    &    $d$ [10$^{-8}$cm]  & $\varepsilon$ [10$^{-24}$cm$^3$]  &
$d/\varepsilon$ ($E_{c}$)  &   $T_{c}$ [K]\\
\hline
Be$^{2+}$ &    0.35 &    (0.008)      &        (43.75)    &
$\simeq 0$ ($<5$) ~\cite{felner01}\\
Mg$^{2+}$ &    0.65 &    (0.094)      &        (6.91)     &
     39 ~\cite{nagamatsu01}\\
Ca$^{2+}$ &    0.99 &    1.1 (0.47)  &      0.9  (2.11) \\
Sr$^{2+}$ &    1.13 &    1.6 (0.86)  &      0.71 (1.31) \\
Ba$^{2+}$ &    1.35 &    2.5 (1.55)  &      0.54 (0.87) \\
Zn$^{2+}$ &    0.74 &    0.8      &         0.93 \\
Cd$^{2+}$ &    0.97 &    1.8      &         0.54 \\
Cu$^{2+}$ &    0.72 &    0.2      &         3.60 \\
Pb$^{2+}$ &    0.84 &    4.9      &         0.17 \\
%
\end{tabular}
\label{table:Ec}
\end{table}
%
%
%
%
%
%
%


\begin{references}

\bibitem{nagamatsu01}
J. Nagamatsu, N. Nakagawa, T. Muranaka, Y. Zenitani and J. Akimitsu:
Nature {\bf 410} (2001) 63.

\bibitem{bianconi01}
A. Bianconi, N.L. Saini, D. Di Castro, S. Agrestini, G. Campi,
A. Saccone, S. De Negri, M. Giovannini and M. Colapietro:
cond-mat/0102410.

\bibitem{kortus01}
J. Kortus, I. I. Mazin, K. D. Belashchenko, V. P. Antropov and L. L. Boyer:
Phys. Rev. Lett. {\bf 86} (2001) 4656.

\bibitem{an01}
J. M. An and W. E. Pickett:
Phys. Rev. Lett. {\bf 86} (2001) 4366.

\bibitem{satta01}
G. Satta, G. Profeta, F. Bernardini, A. Continenza and S. Massidda:
cond-mat/0102358.

\bibitem{mehl01}
M. J. Mehl, D. A. Papaconstantopoulos and D. J. Singh:
cond-mat/0104548.

\bibitem{jin01}
R. Jin, M. Paranthaman, H. Y. Zhai, H. M. Christen, D. K. Christen and
D. Mandrus:
cond-mat/0104411.

\bibitem{hirsch01}
J. E. Hirsch:
Phys. Lett. A {\bf 282} (2001) 392.

\bibitem{felner01}
I. Felner:
cond-mat/0102508.

\bibitem{kawabata94a}
C. Kawabata, S. R. Shenoy and A. R. Bishop:
in {\it Bulletin of the Electrotechnical laboratory} Vol.58, No.6
(ETL, Ibaraki, 1994) p.426(42).

\bibitem{kawabata95a}
C. Kawabata, S. R. Shenoy and A. R. Bishop:
in {\it ADVANCES IN SCIENCE AND TECHNOLOGY 8,
Superconductivity and Superconducting Materials Technologies},
edited by P. Vincenzini
(Techna Srl., Faenza, 1995) p.13.

\bibitem{kawabata95b}
C. Kawabata:
in {\it Advances in Superconductivity VII},
edited by K. Yamafuji and T. Morishita
(Springer-Verlag, Tokyo, 1995) p.233.

\bibitem{kawabata94b}
C. Kawabata, S. R. Shenoy and A. R. Bishop:
in {\it Advances in Superconductivity VI},
edited by T. Fujita and Y. Shiohara
(Spring-Verlag, Tokyo, 1994) p.55.

\bibitem{kawabata96}
C. Kawabata, M. Takeuchi, S. R. Shenoy and A. R. Bishop:
in {\it Advances in Superconductivity VIII},
edited by H. Hayakawa and Y. Enomoto
(Spring-Verlag, Tokyo, 1996) p.271.

\bibitem{kittel}
C. Kittel:
{\it Introduction to Solid State Physics}
(Wiley \& Sons, New York, 1996).

\bibitem{tessman53}
J. R. Tessman, A. H. Kahn and W. Shockley:
Phys. Rev. {\bf 92} (1953) 890.

\bibitem{pauling27}
L. Pauling:
Proc. Roy. Soc. (London) {\bf A114} (1927) 181.

\bibitem{pearson}
W. B. Pearson:
{\it Chemistry and Physics of Metals and Alloys}
(Wiley \& Sons, New York, 1972).



\end{references}
\end{document}